\documentclass[12pt,preprint]{aastex}

\shorttitle{Stable 1:2 Resonance Periodic Orbits}
\shortauthors{Haghighipour et al.}

\begin{document}

\title{Stable 1:2 Resonant Periodic Orbits in
       Elliptic Three-Body Systems}

\author{Nader Haghighipour\altaffilmark{1,2}, 
Jocelyn Couetdic\altaffilmark{3,4}, Ferenc Varadi\altaffilmark{4}
and  William B. Moore\altaffilmark{4,5}}
\altaffiltext{1}{Dearborn Observatory, 
     Northwestern University, Evanston, IL 60208.}
\altaffiltext{2}{Department of Terrestrial Magnetism and the 
NASA Astrobiology Institute, Carnegie Institution of Washington,
5241 Broad Branch Road, Washington, DC 20015;
nader@dtm.ciw.edu.}
\altaffiltext{3}{Option Physique Appliqu\'ee, Ecole Centrale Paris,
        Grande voie des vignes, 92295 Chatenay-Malabry Cedex, France;
couetdj3@cti.ecp.fr.}
\altaffiltext{4}{Institute of Geophysics and Planetary
Physics  and the NASA Astrobiology Institute,
UCLA, 405 Hilgard Avenue, Los Angeles, CA 90095-1567;
varadi@ucla.edu.}
\altaffiltext{5}{Department of Earth and Space Sciences,
UCLA, 595 Charles Young Dr., Los Angeles, CA 90095-1567;
bmoore@avalon.ess.ucla.edu.}

\begin{abstract}
The results of an extensive numerical study of the periodic
orbits of planar, elliptic restricted three-body planetary systems
consisting of a star, an inner massive 
planet and an outer mass-less body in
the external 1:2 mean-motion resonance are presented. 
Using the method of differential continuation,
the locations of the resonant periodic orbits 
of such systems are identified and through an extensive study 
of their phase-parameter space, it is found
that the majority of the resonant 
periodic orbits are unstable.
For certain values of the mass and
the orbital eccentricity of the inner planet, however,
stable periodic orbits can be found.
The applicability of such studies
to the 1:2 resonance of the extrasolar
planetary system GJ876 is also discussed.
\end{abstract}


\section{Introduction}

It is well known that orbital or mean-motion resonances are
of great importance in the dynamics of planetary and satellite systems. 
Resonances drive the dynamical evolution of such systems by 
creating regions of stability and instability in 
orbital phase space. Mean-motion resonances are usually associated with
orbits whose geometrical configurations are periodically repeated.
Such orbits, known as
resonant periodic orbits (hereafter RPOs), are of particular interest
since they define the structure of the associated resonance.
The stability of an RPO implies that it may be a 
potential location for harboring an actual body.
Unstable RPOs, on the other hand,
provide pathways for dramatic orbital evolution of bodies 
and can also be used to explain chaotic motions
\citep{Var99-I}.

Until a few years ago, mean-motion resonances
were found only in our solar system. Recent
success of the precise radial velocity searches 
in detecting more than 100 
extrasolar planets has extended the applicability of resonances
to a broader context, several parsecs beyond the boundaries
of our solar system.\footnote{See exoplanets.org for a complete 
and up-to-date list of extarsolar planets with their 
corresponding references.}
Among these planetary systems, GJ876
is a three-body system with two planets
in a 1:2 resonance \citep{Mar01} and 47 UMa
has two planets that are a near 2:5 
commensurability \citep{Hag03}.
Because of the universal importance and applicability
of resonances, it is of great
value to study RPOs and investigate 
their dynamical stability.

The most widely known RPOs are
the ones associated with the 1:1 resonance. In a three-body
system, these RPOs
are the Lagrangian points in a frame co-rotating with the 
major planet. In case of the three-body system of Sun, Jupiter
and a test particle, the stable RPOs are located at 
$\pm{60^\circ}$ with respect to the location
of Jupiter on its orbit around the Sun. 
The roughly 1500 known Jupiter Trojan asteroids
are the actual bodies that are
trapped in the 1:1 resonance with Jupiter at these two locations.

Our knowledge of the RPOs of other orbital resonances 
is limited to only a few values of 
the eccentricity of the inner planet and its mass
relative to the star. It is therefore
essential to carry out studies similar to those for
the 1:1 resonance for other mean-motion resonances
and to explore their phase-parameter spaces
for regions of stability.
Such a study has been previously carried out by
\citet{Var99-II} for the case of the 2:3 exterior resonance.
In that paper, the method of
differential continuation was employed to identify 
the regions of the phase-parameter space 
of planar, elliptic restricted three-body systems
that correspond to stable orbits of the outer body 
while in the 2:3 resonance. 

In this paper, we present the results of an
extensive study of the RPOs of the exterior 1:2
resonance. The system of interest 
is planar, elliptic and restricted,
with an inner massive planet (hereafter, the major planet)
and an outer zero-mass particle.  
We employ the methodology of \citet{Var99-II} and vary the parameters
of the system in a systematic way in order to identify
its stable RPOs and their corresponding
regions of stability in phase-parameter space.

A review of our methodology is presented in $\S$ 2.
Section 3 presents the results of our numerical
study and in $\S$ 4 we conclude by
reviewing the results and discussing their 
implications to extrasolar planetary systems.

\section{Methodology}

Our method of identifying an RPO consists
of two steps; search and continuation.
To describe this process, we start by briefly reviewing
the condition for an orbit to be periodic.
Let us consider a planar, restricted three-body system with 
a mass-less particle in an exterior resonance with a major planet. If
${\bf x}$ denotes the four-dimensional vector of the initial position
and velocity of the outer particle, then after a certain number
of orbits of the major planet, the particle's position
and velocity will be given by another vector ${\bf F(x)}$. 
The orbit is periodic if
\begin{equation}
{\bf F}\bigl({\bf x}(\varepsilon),\varepsilon\bigr)\,-\,
{\bf x}(\varepsilon)\,=\,0\,,
\end{equation}
\noindent
where $\varepsilon$ is a parameter of the system. 
For the systems considered here,
$\varepsilon$ can be taken to be the ratio of the mass of the
major planet to that of the central star.
The mapping ${\bf F}$ is given by numerically integrating
the equations of motions. Solution vectors ${\bf x}$ of
equation (1) represent RPOs of the system.

Equation (1) can be differentiated with respect to 
$\varepsilon$ to obtain
\begin{equation}
{{d{\bf x}}\over {d\varepsilon}} = 
-{\Bigl({{d{\bf F}}\over {d{\bf x}}}\,-\,{\bf I}\Bigr)^{-1}}\,
{{d{\bf F}}\over {d\varepsilon}}\>,
\end{equation}
\noindent
where ${\bf I}$ is the identity matrix.
Equation (2) involves the variational equations of the dynamics
of the system. It leads to a complicated system of differential equations
in which the independent variable is $\varepsilon$ \citep{Var99-II}.

Equation (1) can, in principle, be solved by Newton's method. 
Because the latter converges only for some vectors
${\bf x}$, due to the nonlinear nature of equation (1),
many initial vectors have to be tried.
For a chosen value of $\varepsilon$, 
we select random initial conditions,
integrate the dynamical equations and if ${\bf F}$ happens to be
small, we apply Newton's method to search for a nearby solution
of equation (1). 
We then apply differential continuation \citep{Kel77}, 
i.e., solve equation (2), to follow the changes in the RPO due to 
varying $\varepsilon$.

We start our search for 1:2 resonant periodic orbits by
choosing the semimajor axis of the outer body near its
resonant value with the inner planet. 
The semimajor axis of the inner planet is assumed to be
equal to unity. Many possible 
configurations are considered in which the other orbital
parameters of the outer body are chosen randomly.
The inner planet is placed at either the peri- or the apocenter
without any loss of generality. We integrate the system
for two orbits of the inner planet and form the differences 
between the initial and the final position and velocity vectors 
of the outer body. When these differences are small, we apply
Newton's method to further reduce them.
This process is repeated until these differences become
so small that the orbit  can be considered to be an RPO.

The ratio of the mass of the inner planet to the mass of the
central star $(\mu)$ and also the orbital eccentricity of the
inner planet $(e)$ are the parameters of our system. During the
process of differential continuation, we vary $\mu$ and $e$
and solve equation (2) to obtain new RPOs along a path
in $(\mu, e)$ space. To identify the regions of the phase-parameter 
space of the system which correspond to
stable RPOs, we change $\mu$ and $e$ systematically and
criss-cross the $(\mu, e)$ space with
paths of continuation along which the stability of
RPOs is computed.

The stability of RPOs is examined by computing the eigenvalues
of the monodromy matrix $d{\bf F}/d{\bf x}$ \citep{Dan64,Gol96}. 
Stable RPOs are centers of libration
for nearby orbits. Unstable RPOs are associated with
separatrix-like features \citep{Guc83}. 
The eigenvalues of the monodromy matrix provide
the frequency of libration associated with stable RPOs
and also the Lyapunov exponent of the
outer body when the RPO is unstable. 
The system is linearly (spectrally) stable if these 
eigenvalues are on the unit circle in the complex plane.
There is usually a separation of time scales between fast
and slow dynamics and both can correspond to either 
stable or unstable motions. 
The fast dynamics usually consists of motions which are
also present
in the case of a circular orbit for the major planet.
When this fast motion is stable, we call it libration.
The slow dynamics is associated with the effects of orbital
eccentricity of the major planet. We call stable motions in this case
secular librations and unstable ones secular instabilities.
An RPO is unstable when either of the two motions is unstable.

\section{Numerical Results}

We studied the orbit of the outer body of our three-body system
for different values of the
mass-ratio $\mu$ and also different values of the
orbital eccentricity of the inner
planet, $e$. To describe the shape and the orientation of these
orbits, we use 
\begin{eqnarray}
&h=e\cos \varpi\,,\qquad k=e\sin\varpi \\
&h'=e'\cos \varpi'\,,\quad k'=e'\sin\varpi',
\end{eqnarray}
\noindent
where $e'$ is the orbital eccentricity
of the outer body and $\varpi$ and $\varpi'$ represent the
longitudes of the pericenters of the inner planet and the
outer body, respectively. 
In order to completely specify an RPO, 
one needs the $h'$, $k'$, semi-major axis and mean longitude
of the RPO at a given orbital phase of the inner planet. 

We performed an extensive initial numerical search for 1:2 resonant
periodic orbits for several mass-ratios and 
orbital eccentricities of the inner planet. 
Once an RPO
was found, we used its orbital parameters for starting the
continuation process. Figure 1 shows the overall results
of the continuation process for starting values of 0.001 and 
0.1 for $\mu$ and $e$, respectively. For all the shown branches
of continuation, we have we have $\varpi$ = 0 and therefore 
$k=0$ and $h=e$. For the same orbit of the
inner planet, there can be several RPOs with different 
orbital elements on different branches.
These branches are labeled according to the initial
values of their $h'$ and $k'$ quantities.
This method of labeling enables
one to readily calculate the orbital eccentricity and the
longitude of the pericenter of the outer body at the start of
the continuation process.
For instance, the branch labeled as $h'52k'16$ 
represents the result of continuation started from
a resonant periodic orbit with $h'=0.52$ and $k'=0.16$. Such an orbit has
an eccentricity equal to 0.544 and a longitude of
pericenter equal to 17.1$^\circ$. 

One can see from Figure 1 that certain branches shown here
appear to be connected, either through so-called
turning points or via circular connection.
To explain these cases,
we recall that any continuation branch corresponds
to a periodic orbit that has been continued from an initial
RPO, for different values of the mass
and the orbital eccentricity of the inner body. 
The continuation process can break down at points where the matrix 
$ (d{\bf F}/d{\bf x}-{\bf I}) $
becomes singular. 
When two branches of continuation meet at a point
on the $h'$-$h$ plane where $h\ne 0$, they are said to be connected
through a turning point. Turning point is a general term
indicating that a branch changes its direction, or in other word,
at a turning point, the tangent line to the continuation
branch will be vertical (e.g., \citealt{Kel77}). 
The two branches $h'15k'00$ and
$h'59k'00$ are connected via a turning point. Figure 2 shows
the changes of the different orbital parameters of the outer body
for these two branches. 

On the line $h=0$, the matrix above is always singular due
to the fact that the differential equations for the outer
planet become time-independent.
In this case, the variational equations transport the
equations of motions invariantly along the orbit
(e.g., \citealp{Sie71,Var99-II}).
A circular connection occurs when two
branches of continuation meet on the $h=0$ line.
Branches $h'004k'00$ and $h'15k'00$, $h'59k'00$ and $h'-$67$k'00$,
and also $h'-$05$k'-$57 and $h'52k'16$ are connected in this way.
We also have to note that the matrix above can also become
singular at bifurcation points, which is discussed in the
context of stability analysis.

The continuation process, however, can sometimes pass through
singularities without major difficulties. This is 
due to the fact that the particular singularities are
handled gracefully by the numerical integrator used to solve
equation (2). An example is the smooth connection
between $h'004k'00$ and $h'15k'00$ in Figure 1.
Eventually there is  a singularity of some type through which 
continuation is not possible. 

As mentioned above, periodic orbits of the outer body
are labeled by their orbital elements
at the pericenter passage of the inner planet. In the cases
in which the continuation process can be extended beyond $h=0$,
i.e., beyond the orbit of the inner planet being circular,
the pericenter and the apocenter of the inner planet exchange
places. That is, pericenter becomes apocenter and vice versa.
To assure that the orbital parameters of the outer body
will still be calculated at the time of the pericenter
passage of the inner planet, we integrate the equations of 
motion of the outer body for half of the orbital period of
the inner planet and then rotate the coordinate systems in
such a way that the pericenter of the inner planet is located in
the positive direction of the $x$-axis. At the same time,
since we are dealing with the 1:2 external resonance,
the outer body travels only one-fourth of its orbit and
one obtains a new set of orbital elements.
Once the re-alignment process is completed, there will be two sets 
of orbital parameters associated with the outer body. We
assure that these two sets describe 
orbits unambiguously  by limiting
the values of the mean longitude of the outer body, $\lambda'$. 
In this study ${-90^{\circ}}<{\lambda'}<{90^{\circ}}$. 
The re-alignment correction becomes necessary when the orbital
parameters of the outer body undergo sudden changes during the
passage of the inner planet through a circular orbit. Figure
3 shows such changes for the re-aligned and non-realigned values
of $h$ and $h'$ for two circularly connected branches of
$h'004k'00$ and $h'15k'00$. 

As mentioned earlier, 
Figure 1 provides an overall view of the different continuation
branches for a specific value of the inner
planet's mass. When the latter is also varied, the branches
become leaves parameterized by the mass and the eccentricity of the
inner planet. In order to explore these leaves, the 
$(\mu, e)$ space is criss-crossed with continuation paths.
Along these paths, $\mu$ and $e$ are assumed to depend on a common
independent variable. The derivatives of $\mu$ and $e$ with
respect to the independent variable are prescribed to yield
the desired continuation path. Since the actual variable used
in the continuation is not $e$ but the velocity of the inner planet 
at peri- or apocenter, the paths will have slight curvature 
(see also \citealp{Var99-II}). 

We also studied the phase-parameter space of the system, 
in search of regions corresponding to stable resonant
periodic orbits, for all different
branches of Figure 1. In most cases, the RPOs were unstable.
Figure 4 shows a criss-crossing of the $(\mu, e)$ space with
continuation paths of the RPO designated as
$h'004k'00$. An interesting feature of this 
case is that the RPO can be stable in the horizontal 
directions while unstable in the vertical one.
We note that planar motions, such as the RPOs in this
study, can be unstable to vertical perturbations.
Straightforward algebra reveals that the variational
equations decouple the linearized dynamics of 
horizontal and vertical motions. Hence it is
relatively simple to analyze vertical stability.

In Figure 5 and others of the same type, the quantities
on the vertical axes represent either frequencies of
libration or Lyapunov exponents, both computed from the
eigenvalues of the monodromy matrix. When the plotted
value is positive, it is known as the libration frequency.
When the plotted values is negative, its absolute value
is conventionally called Lyapunov exponent.
Since both quantities have units of time, it is possible
to plot them on the same axis. \citet{Var99-II} provides
some clarifying examples.
The orbit is linearly stable if all plotted
values are positive and it is unstable otherwise.
Since the linearized equations are uncoupled for
horizontal and vertical motions, there is a single frequency
of libration or Lyapunov exponent associated with the
vertical motions. The horizontal motions are associated with
a pair of quantities, each being either frequency of libration
or Lyapunov exponent.

In Figure 5, we have plotted frequencies of librations or
Lyapunov exponents along the vertical continuation
path $e=0.5$ in Figure 4. For the values of the mass-ratio, $\mu$,
considered in Figure 4, the orbit is unstable.
From Figure 5a, one
can see that  while this instability is sustained
along the vertical direction and for 
all values of $\mu$, the orbit becomes horizontally stable when
$0.0091 \leq \mu \leq 0.03$ and $0.099 \leq \mu \leq 0.115$.

Figure 6 shows similar criss-crossing analysis for the
continuation branch of the RPO $h'$-05$k'$-57. The system, in this case,
reveals a region of stability with a distinct
boundary from its large region of instability.
The frequencies of librations or Lyapunov exponents
of the system for different
values of the mass-ratio $\mu$ and the orbital eccentricity of the
inner planet are shown in Figures 7, 8 and 9. 
As shown in Figure 7, the orbit of the outer planet is stable
along the horizontal line of $\mu = 0.001$ in Figure 6, for all
values of the orbital eccentricity of the outer planet less than
0.57. Along a vertical line such as $e=0.5$ in Figure 6, 
the system is stable for all values of the mass-ratio less than
0.0014 (Figure 8). Figure 9 shows frequencies of librations or 
Lyapunov exponents of the
orbit of the outer body for the continuation branch 
in which the orbital eccentricity and the mass-ratio of the
inner planet increase in steps of 0.01 and 0.001, respectively.
The continuation path corresponding to this figure has been
denoted by {\bf X} in Figure 6. As shown in Figure 9, the
orbit of the outer planet is stable for the values of the
mass-ratio $\mu$ less than 0.005 and it becomes unstable
for its larger values.

Figure 6 also shows that at higher values of the eccentricity of
the inner planet, the stability of the system requires lower
mass ratios implying that the RPOs of a system with a more
massive inner planet in a 1:2 exterior resonance are most
stable when the orbit of the inner planet is closer to a
circular orbit. The actual boundaries of stable regions
can be different for different resonances and different
leaves of the same resonance. For instance, the case
of the 2:3 external resonance has a small wedge of
instability \citep{Var99-II}.

Figures 10a and b show the different
RPOs of such a system for the mass-ratio of 0.001
and different values of the inner planet's orbital eccentricity.
It is worth noting that for these stable orbits, the pericenter
directions of the inner and outer planets are not aligned.
The existence of a similar asymmetry of stable RPOs 
has also been suggested in the context of the Galilean 
satellites \citep{Gre87}.

\section{Discussion}

We presented the results of a systematic search for stable
resonant periodic orbits of planar, elliptic and restricted 
three-body systems in the exterior 1:2 resonance.
Our approach in this paper toward the understanding of orbital resonances 
provides fundamental information on their structures.
Stable periodic orbits are centers of librations whose
periods are also obtained, at least for small amplitudes,
from our stability computations. 

Using the method of differential continuation, we examined the
phase-parameter space of the system for different values of the mass-ratio
and the orbital eccentricity of the inner planet.
In the majority of the cases, the RPOs were unstable. 
We were, however, able to show that stable resonant periodic orbits 
can be found when the mass and the orbital eccentricity of
the inner planet are within a certain range (Figure 6). 

Our initial search for RPOs also covered the internal 2:1 resonance.
We have not found any RPOs, either stable or unstable, in this case.
This can have significant implications 
for the 2:1 Kirkwood gap in the asteroid belt.
It might be possible to arrive at an explanation of 
the gap in terms of the fundamental properties of the Sun-asteroid-Jupiter
three-body system. This approach would be an
alternative to the one adopted by \citet{Moo98}
which employ elaborate physical models.

The apparent lack of RPOs in the internal 2:1 resonance 
also raises questions regarding the same resonance in the full
three-body system, i.e., when the mass of the outer body is non-zero.
As the latter increases and the mass of the inner body
decreases, the system passes from external resonance
to internal. Along the way, one should find the limits
of masses for which stable RPOs exist. 
Such studies will have immediate applications to some of
the recently discovered extrasolar planetary systems such
as GJ876 in which two planets are locked in a near 1:2
resonance.

\acknowledgments
We benefited from stimulating conversations with S. Peale.
This work was partially supported by
NASA Astrobiology Institute under grants NCC2-1056 (N.H.)
and NCC2-1050 (J.C. and F.V.) and also by NASA NAG grant 5-11691
(F.V. and W.B.M.).

\clearpage

\begin{figure}
\plotone{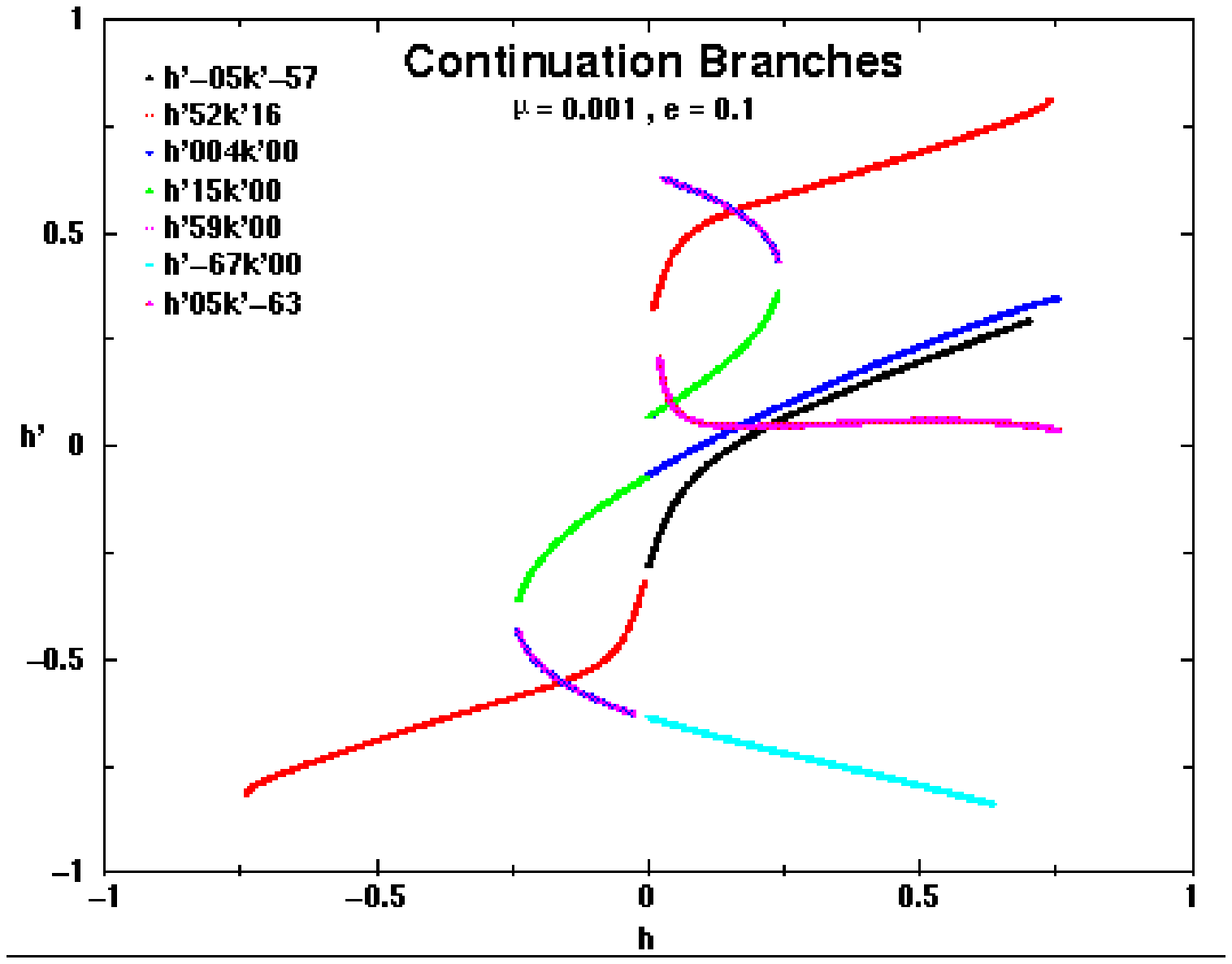}
\caption{
Structure of the continuation branches for
different values of the orbital eccentricity of the outer
body. Each branch represents the results of the continuation
process started from an RPO with the given values of $h'$
and $k'$. The mass-ratio and the orbital eccentricity of the inner planet
at the start of the continuation process were
$\mu$=0.001 and $e$=0.1, respectively.
\label{fig1}}
\end{figure}

\clearpage 
\begin{figure}
\plotone{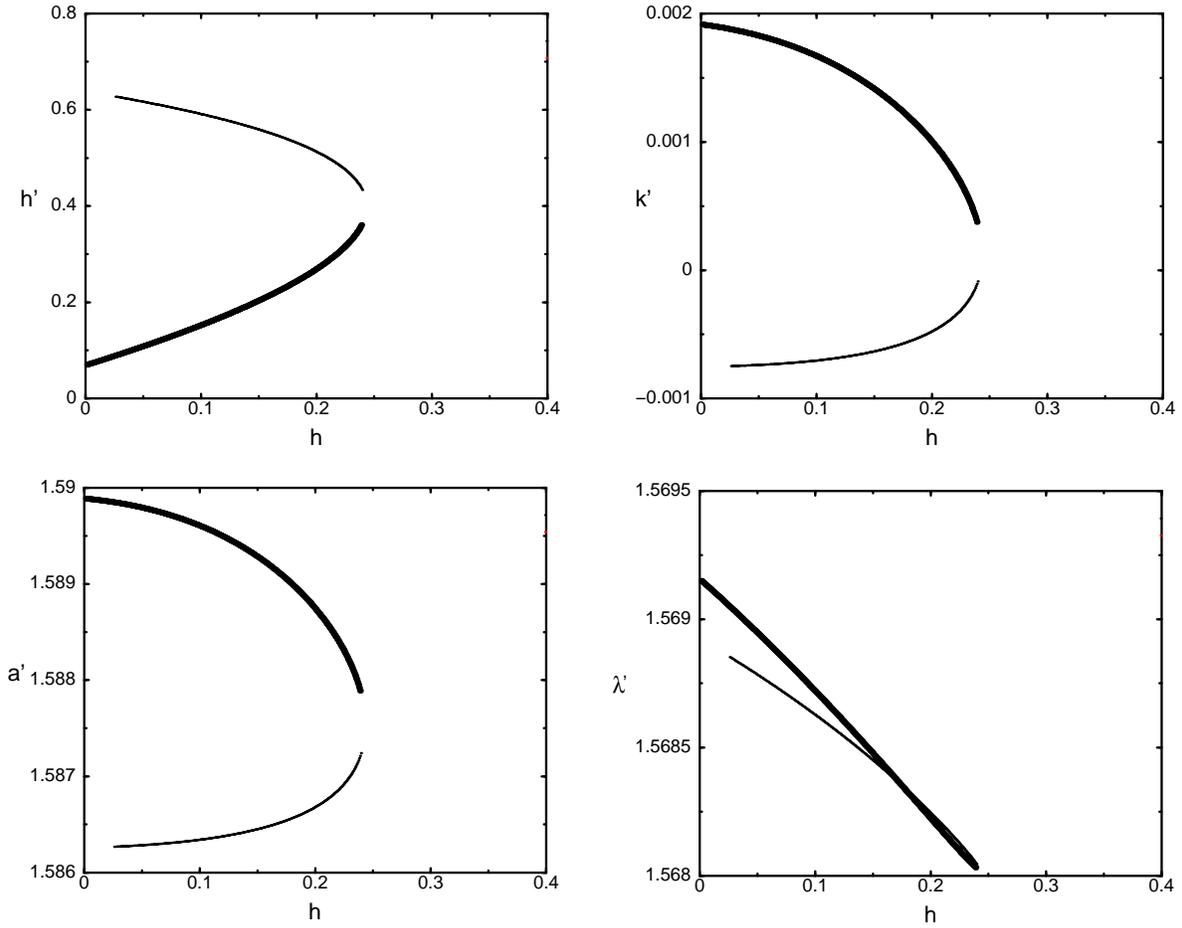}
\vskip -2.5in
\caption{
Graphs of the orbital parameters of the outer body versus
$h$ for turning-points connected continuation branches
$h'$15$k'$00 (thick line) and $h'$59$k'$00 (thin line). 
The quantity $\lambda'$ is the mean longitude of the outer body. 
The starting values of $\mu$ and $e$
for all these graphs are equal to 0.001 and 0.1, respectively.
\label{fig2}}
\end{figure}

\clearpage 
\begin{figure}
\plotone{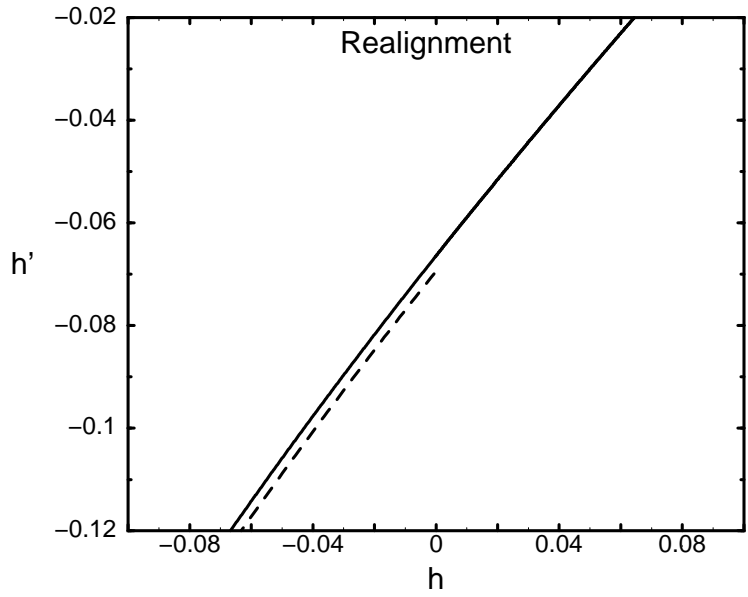}
\vskip -2.5in
\caption{
Graph of $h'$ versus $h$ for two circularly connected branches
$h'$004$k'$00 and $h'$15$k'$00 before
re-alignment (solid line) and after
re-alignment (dashed line).
\label{fig3}}
\end{figure}

\clearpage 
\begin{figure}
\plotone{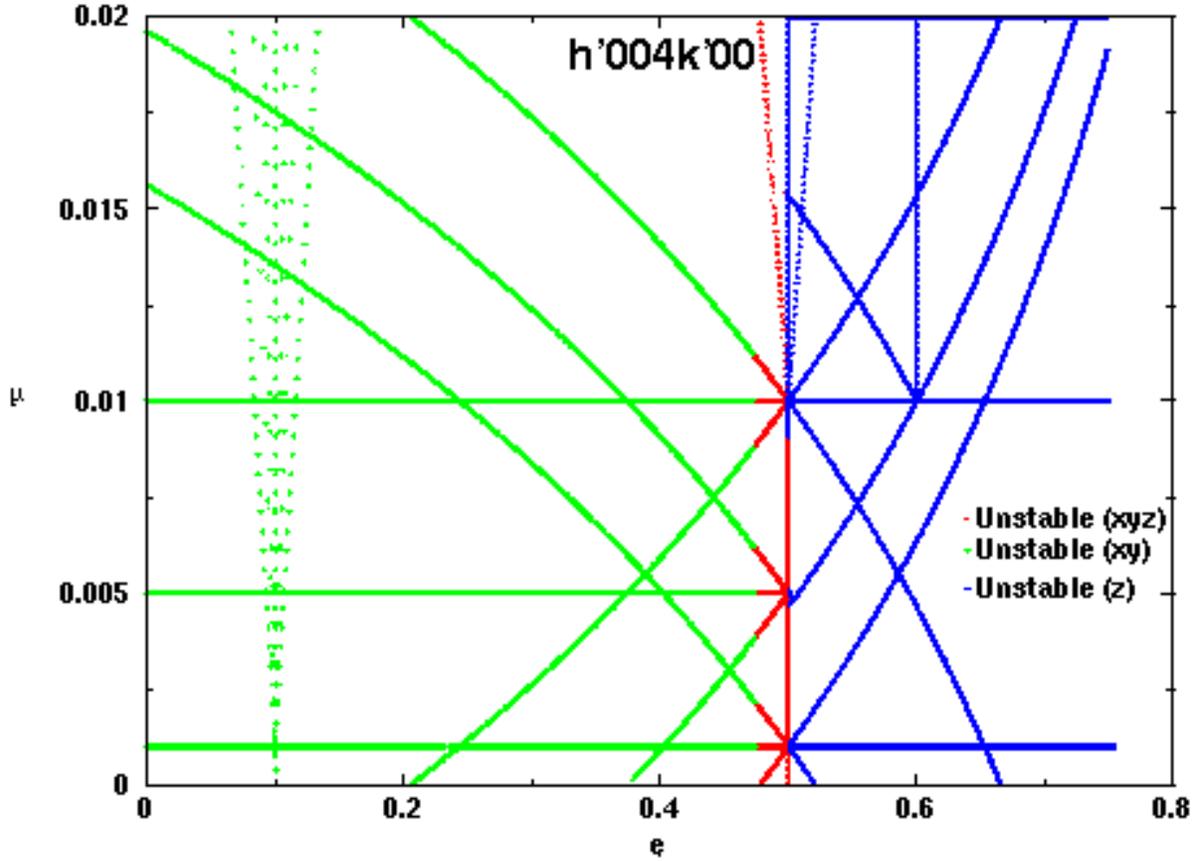}
\caption{
Continuation branches criss-crossing the 
phase-parameter space
of the system. The starting RPO for all these branches
is $h'$004$k'$00. This figure shows that in this case, 
for the chosen values of the mass-ratio and orbital
eccentricity of the inner planet, there is no region 
of the parameter space which corresponds to stable resonant
periodic orbits.
\label{fig4}}
\end{figure}

\clearpage 
\begin{figure}
\plotone{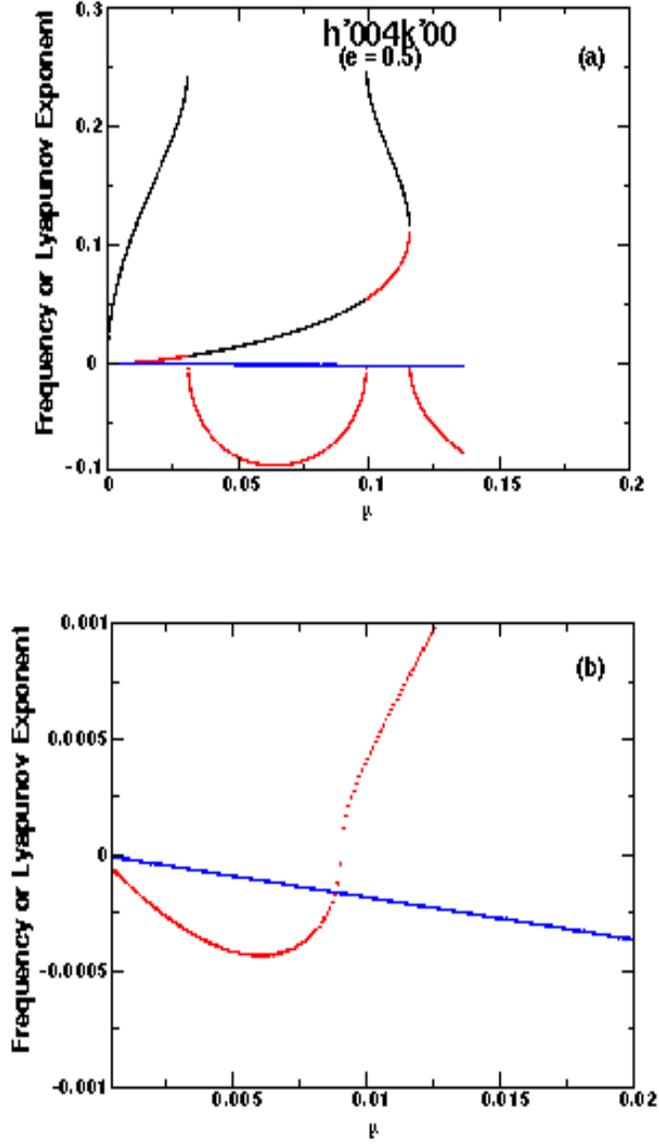}
\vskip -2.5in
\caption{
Frequency of libration (black)and secular libration (red)
on the $xy$ plane and the Lyapunov exponent along the
$z$-axis (blue) of the outer body while the eccentricity of
the inner planet is constant at 0.5. 
Figure 5b shows that for the values of the mass-ratio
indicated in Figure 4, the system is 
unstable to both horizontal and vertical perturbations.
While the system sustains its instability along the vertical
direction, it becomes stable horizontally for 
$0.0091 \leq \mu \leq 0.03$ and $0.099 \leq \mu \leq 0.115$.
\label{fig5}}
\end{figure}

\clearpage 
\begin{figure}
\plotone{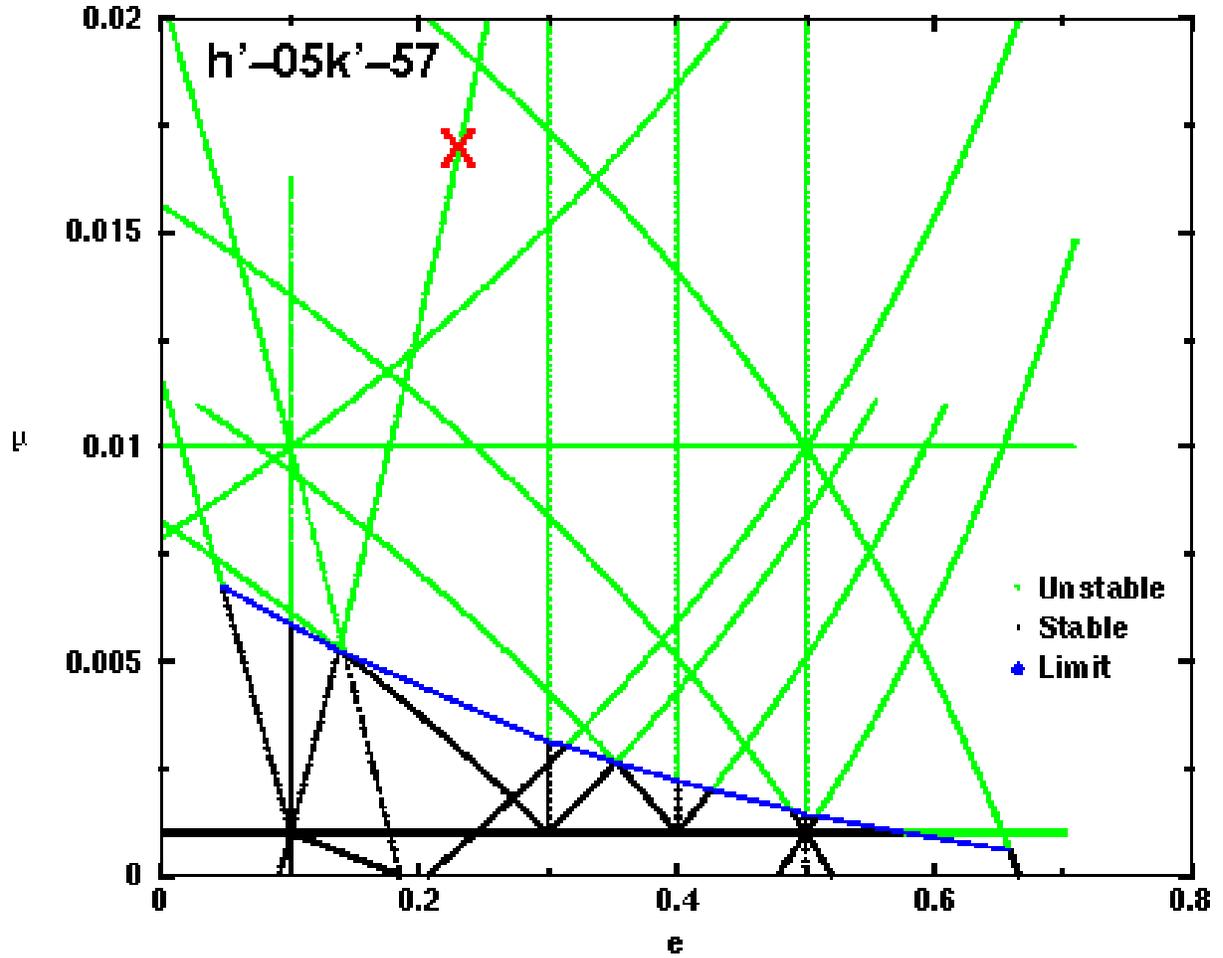}
\caption{
Criss-crossing the phase-parameter space of the system
by continuation branches corresponding to the RPO,  
$h'-$05$k'-$57. One can see that in this case, there
is a distinct region of the phase-parameter space where
the resonant periodic orbits are stable.
\label{fig6}}
\end{figure}

\noindent
\clearpage 
\begin{figure}
\plotone{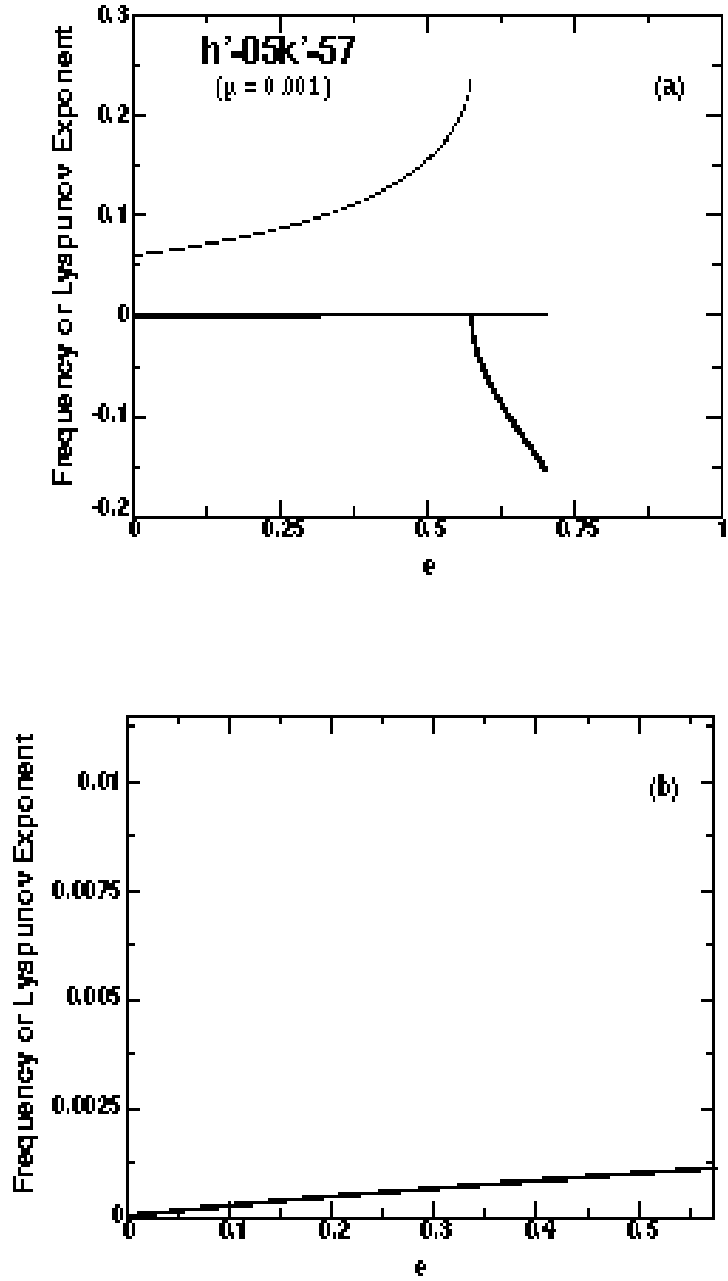}
\vskip -2.5in
\caption{
Frequency of libration (thin line) 
and secular libration (thick line)
for the system of Figure 6 along the horizontal line of $\mu = 0.001$.
As shown in Figure 7b, the orbit of the outer body is stable
for the values of the orbital eccentricity of the outer planet
smaller than 0.576.
\label{fig7}}
\end{figure}

\noindent
\clearpage 
\begin{figure}
\plotone{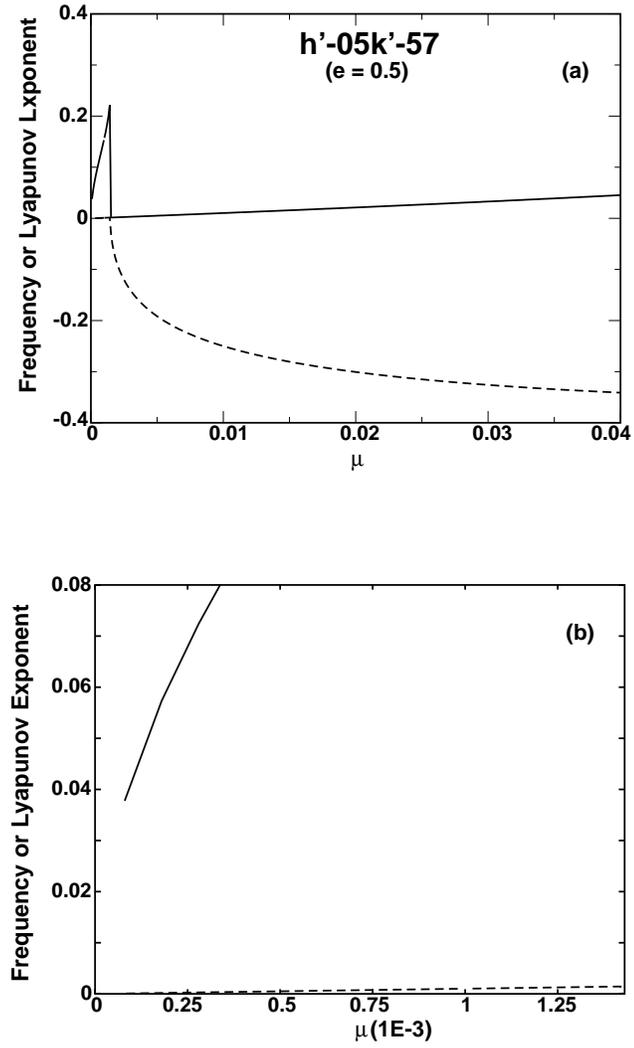}
\vskip -2.5in
\caption{
Frequencies of libration (solid line) and secular libration
(dashed line) for the vertical line of $e=0.5$ in Figure 6.
As shown in Figure 8b, the orbit of the outer planet is stable
for the values of the mass-ratio $\mu$ less than 0.0014.
\label{fig8}}
\end{figure}

\noindent
\clearpage 
\begin{figure}
\plotone{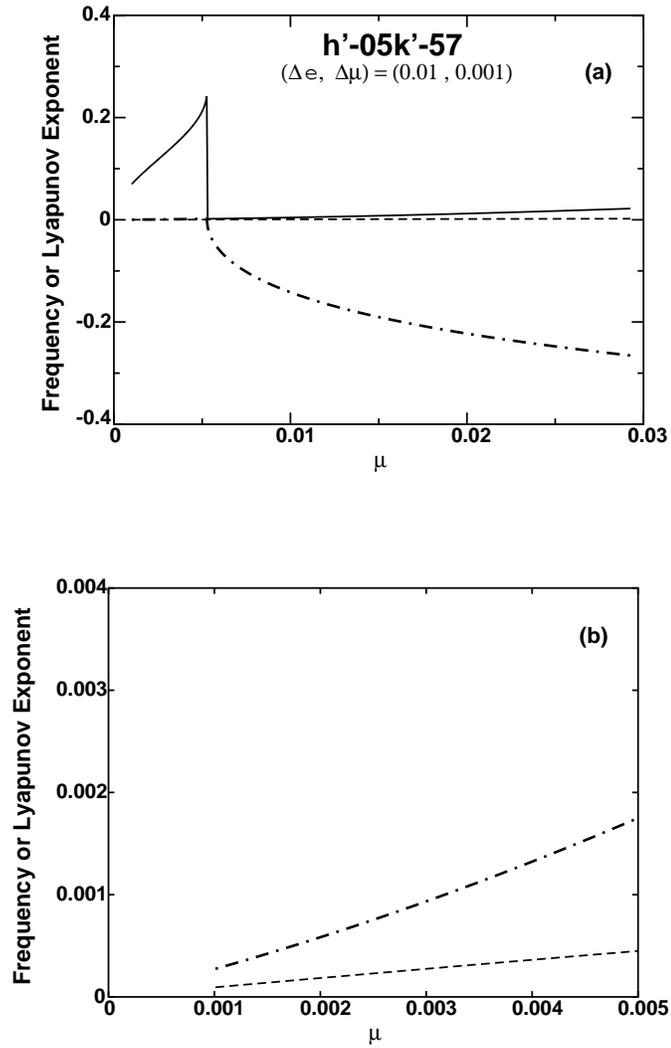}
\vskip -2.5in
\caption{
Frequencies of libration (solid line) and secular libration 
(dash-dotted line) and the Lyapunov exponent 
(dashed line) of the orbit of the outer body for the continuation
path indicated by {\bf X} in Figure 6. As shown in Figure 9b,
the orbit of the outer planet is stable for the values of the
mass-ratio less than 0.005.
\label{fig9}}
\end{figure}

\clearpage 
\begin{figure}
\plotone{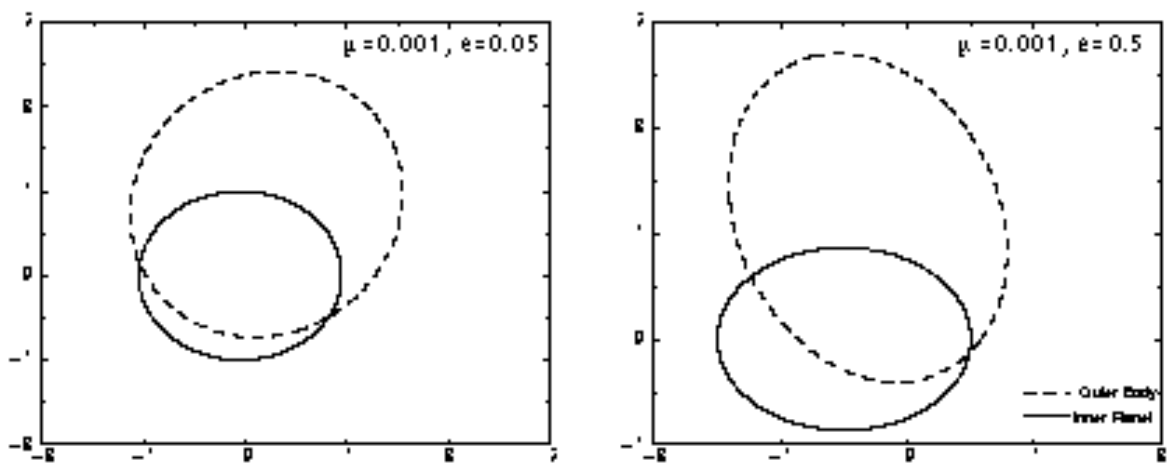}
\caption{
Resonant periodic orbits for $h'$-05$k'$-57.
\label{fig10}}
\end{figure}

\end{document}